\documentclass[preprintnumbers,11pt,a4paper,oneside]{article}
\pdfoutput=1

\usepackage{contour}
\usepackage{xcolor}

\usepackage{mathrsfs}

\contourlength{0.05em} 

\usepackage{jheppub}
\usepackage{color}
\usepackage{graphicx}
\usepackage{wrapfig,enumerate,slashed}

\usepackage{footmisc}
\usepackage{amsmath}
\usepackage{wasysym} 
\usepackage{graphicx}
\usepackage{subcaption}
\usepackage{color}
\usepackage{comment}
\usepackage{appendix}
\usepackage{slashed}
\usepackage{siunitx}

\newcommand{\ket}[1]{\left| #1 \right>} 
\newcommand{\matrixel}[3]{\left< #1 \vphantom{#2#3} \right|
	#2 \left| #3 \vphantom{#1#2} \right>} 

\usepackage{tikz}
\usepackage{orcidlink}

\makeatletter
\gdef\@fpheader{}
\makeatother
\begin{document}

\title{Enhancing Quantum Field Theory Simulations on NISQ Devices with Hamiltonian Truncation}

\author[a]{James~Ingoldby,\orcidlink{0000-0002-4690-3163}}
\author[a]{Michael Spannowsky,\,\orcidlink{0000-0002-8362-0576}}
\author[a]{Timur Sypchenko,\,\orcidlink{0009-0008-3894-7458}}
\author[a]{Simon Williams\,\orcidlink{0000-0001-8540-0780}}

\emailAdd{james.a.ingoldby@durham.ac.uk}
\emailAdd{michael.spannowsky@durham.ac.uk}
\emailAdd{timur.sypchenko@durham.ac.uk}
\emailAdd{simon.j.williams@durham.ac.uk}

\affiliation[a]{\vspace{0.1cm} Institute for Particle Physics Phenomenology, Durham University, Durham DH1 3LE, UK}

\preprintA{IPPP/24/37}

\abstract{Quantum computers can efficiently simulate highly entangled quantum systems, offering a solution to challenges facing classical simulation of Quantum Field Theories (QFTs). This paper presents an alternative to traditional methods for simulating the real-time evolution in QFTs by leveraging Hamiltonian Truncation (HT). As a use case, we study the Schwinger model, systematically reducing the complexity of the Hamiltonian via HT while preserving essential physical properties. For the observables studied in this paper, the HT approach converges quickly with the number of qubits, allowing for the interesting physics processes to be captured without needing many qubits. Identifying the truncated free Hamiltonian's eigenbasis with the quantum device's computational basis avoids the need for complicated and costly state preparation routines, reducing the algorithm's overall circuit depth and required coherence time. As a result, the HT approach to simulating QFTs on a quantum device is well suited to Noisy-Intermediate Scale Quantum (NISQ) devices, which have a limited number of qubits and short coherence times. We validate our approach by running simulations on a NISQ device, showcasing strong agreement with theoretical predictions. We highlight the potential of HT for simulating QFTs on quantum hardware.}

\maketitle


\section{Introduction}

The simulation of Quantum Field Theories (QFTs) is crucial for understanding fundamental particles and their interactions and forms the cornerstone of modern particle theory. Despite their success, simulating QFTs on classical computers is difficult, and contemporary approaches suffer from challenges which severely limit the exploration of QFTs, such as the so-called sign problem, which is a fundamental difficulty in Monte Carlo importance sampling algorithms~\cite{PhysRevLett.46.77, VONDERLINDEN199253, PhysRevE.49.3855}. These challenges are especially pronounced in areas such as Quantum Chromodynamics (QCD) at finite density and non-perturbative calculations of the real-time dynamics of hadrons.

Quantum computers have the ability to simulate the real-time dynamics of highly entangled systems, therefore offering a solution to the sign problem and providing a natural regime for simulating QFTs. As a result, there has been great interest in designing algorithms for the simulation of QFTs on quantum devices~\cite{Banuls:2019bmf, Jordan:2011ci, Jordan:2012xnu, Jordan:2014tma, Araz:2022tbd, Fromm:2023npm, Abel:2024kuv}, with most approaches choosing to adopt the Kogut-Susskind Hamiltonian formulation for SU($N$) Lattice gauge theory~\cite{PhysRevD.11.395}. The Lattice approach, whilst powerful, has its own set of limitations. One of the most prominent problems facing this approach is the initial-state preparation, achieved by preparing specific quantum states on the device using state-preparation routines. These routines are known to be extremely costly, and the resources required to implement the state-preparation for an arbitrary state can scale exponentially~\cite{Xiaoming10044235, PhysRevA.83.032302, PhysRevResearch.3.043200, PhysRevLett.129.230504}. 

This paper presents an alternative method which leverages Hamiltonian Truncation (HT) techniques to facilitate the non-perturbative, real-time simulation of QFTs on a quantum device. It will be shown that this approach allows for the Hamiltonian to be constructed such that complicated and costly state-preparation routines are not needed to simulate the time evolution of a QFT. Furthermore, for the observables studied in this paper, it will be shown that the HT approach converges quickly as the truncation increases, thus allowing for the system's dynamics to be captured reliably without needing many qubits. As a result, the HT approach has a reduced circuit depth on fewer qubits, making the method well suited for Noisy-Intermediate Scale Quantum (NISQ) devices, which have limited numbers of qubits and short coherence times. 

Hamiltonian Truncation is a non-perturbative numerical method for approximating the spectrum and dynamics of strongly coupled quantum systems. In this approach, the Hamiltonian is expressed on a truncated basis of states. It is particularly useful when applied to QFTs because selecting a finite number of basis states reduces the infinite-dimensional Hamiltonian of a QFT to a finite-dimensional one, making computational analysis possible. Furthermore, HT can be applied directly to continuum QFT Hamiltonians without discretising them first on a lattice, removing the need to control systematic uncertainties arising from lattice artifacts. This formalism, which is an alternative to the lattice, has enabled the study of a wide variety of phenomena, including critical points in scalar field theories~\cite{PhysRevD.91.025005,Rychkov:2014eea,Rychkov:2015vap,Elias-Miro:2017xxf,Elias-Miro:2017tup,Anand:2020qnp}, scattering processes \cite{Bajnok:2015bgw,Gabai:2019ryw,Henning:2022xlj}, and the decay of metastable vacua \cite{Szasz-Schagrin:2022wkk,Lencses:2022sfa}. For a general introduction and review of applications, see References~\cite{Konik-review,fitzpatrick2022}. 

Hamiltonian Truncation, therefore, provides a powerful framework to explore and understand systems where traditional perturbative techniques fail and favourably compares to lattice gauge theories in their resource requirements for Hamiltonian simulation. Alternative truncation schemes have been considered in Reference~\cite{Li:2024lrl}.

To explore and illustrate the efficacy of the HT approach and its suitability to NISQ devices, we consider the massive Schwinger model~\cite{Schwinger:1962tp, Coleman:1976uz, 10.1093/oso/9780198834625.001.0001}, which is analogous to Quantum Electrodynamics (QED) in $\left(1+1\right)$ dimensions. The model provides a simple yet rich framework to test the quantum simulation of QFTs via HT. The Schwinger model is exactly solvable in its massless form and offers insights into phenomena such as confinement and screening, which are crucial for understanding more complex theories like QCD. 

In Section~\ref{sec:HT}, the HT procedure is outlined by first bosonising the model, which re-expresses it as an interacting scalar field theory and removes all the gauge redundant degrees of freedom. This scalar field theory is constructed on the finite-volume circle before applying HT, without introducing a lattice. The process of truncating the Hilbert space reduces the complexity of the Hamiltonian whilst ensuring that it retains the essential features and interactions to simulate real-time evolution accurately. Consequently, the truncated Hamiltonian is more suited to implementation on NISQ devices with limited qubits and coherence times. 

In Section~\ref{sec:QC}, we demonstrate this suitability by considering the dynamics of the model after a quench and evaluating time-evolution using different truncation levels. We see that the model gives a good description of the post-quench dynamics of the Schwinger model, even with a small number of resources, ultimately allowing for the model to be run on the \texttt{ibm\_brisbane} quantum computer. Utilising the control and error suppression techniques provided by \textsc{Q-Ctrl}~\cite{boulder_opal1, boulder_opal2} to improve the fidelity of the results, we obtain the first simulation of a QFT on a quantum device through HT. The results show a remarkably accurate simulation of the post-quench dynamics of the Schwinger model on a NISQ device.
 
This work highlights the potential of HT for simulating QFTs on near-term quantum hardware, achieving accurate and precise results with low resource costs. Furthermore, it paves the way for future research in leveraging quantum devices for non-perturbative field theory simulations.


\section{Hamiltonian Truncation}\label{sec:HT}

Hamiltonian Truncation (HT)~\cite{Konik-review,fitzpatrick2022} involves approximating the full Hamiltonian of a theory by restricting it to a finite subspace of the Hilbert space. This technique is particularly useful when perturbative methods fail, such as in strongly coupled systems or when non-perturbative effects are significant.  

In this approach, the first step is to decompose the Hamiltonian of the quantum system of interest into a solvable part $H_0$, plus an interaction $V$ so that
\begin{align}\label{eq:hdef}
	H = H_0 +V\,.
\end{align}
The eigenstates and corresponding eigenvalues of the solvable part can then be identified and labelled $H_0 \ket{i} = E_i \ket{i}$. Next, a truncation scheme is introduced. An energy cutoff, denoted as $E_\textrm{max}$, is applied, and we retain only states with $H_0$ eigenvalues up to this cutoff $E_i \le E_\textrm{max}$ in the truncated Hilbert space. To the extent that low-energy states in the full theory lie within this truncated Hilbert space, the low energy physics of the full theory will be well approximated in HT.

The truncated Hamiltonian is then represented as a matrix, which acts on states of a truncated basis. Numerical diagonalisation can then yield approximate eigenvalues and eigenvectors for the system. The eigenvalues correspond to the approximate energy levels, while the eigenvectors provide information about the corresponding quantum states.

The results are then analysed, checking the convergence with respect to the cutoff  $E_\textrm{max}$. This involves increasing  $E_\textrm{max}$ and stabilising physical observables such as energy levels or correlation functions. The physical interpretation of these results is then made in the original problem's context, often compared with known analytical results or experimental data if available. 

In the QFT context, truncation of the basis acts as a kind of nonlocal UV regularisation,\footnote{The regularisation is nonlocal because the cutoff  $E_\textrm{max}$ acts on the total energy of a state and does not take into account how widely separated in space different particles or excitations in the same state are.} so that in a QFT with UV divergences, there will be observables that diverge as $E_\textrm{max}$ is increased. In this case, nonlocal counterterms must be added to the Hamiltonian to renormalise the theory. A systematic procedure for constructing the nonlocal terms is outlined in References~\cite{EliasMiro:2022pua,Delouche:2023wsl}. Even in UV finite QFTs, many observables converge in a power like fashion (i.e. as $O\sim O_\infty+c/E_\textrm{max}^p$ for a model-dependent positive power $p$) as the cutoff is increased, and it can be beneficial to add improvement terms to the Hamiltonian which improve the rate of convergence~\cite{PhysRevD.91.025005,Rychkov:2014eea,Rychkov:2015vap,Elias-Miro:2015bqk,Elias-Miro:2017xxf,Elias-Miro:2017tup,Cohen:2021erm,Lajer:2023unt}. These terms are constructed to account for states above the HT cutoff on low energy dynamics, and are analogous to higher dimension effective operators included in the action of a low-energy effective field theory.

Hamiltonian truncation is a versatile technique adaptable to various quantum systems and field theories. It has the potential to bridge the gap between exact analytical solutions and purely numerical approaches like lattice field theory, offering a tool for studying complex quantum phenomena.

\subsection{Hamiltonian Truncation applied to the Schwinger Model}

The Schwinger model has become a benchmark scenario for comparing non-perturbative methods in simulating field theories, ranging from lattice gauge theories \cite{PhysRevD.13.1043, PhysRevD.56.55, HAMER1982413, PhysRevD.57.5070, PhysRevResearch.4.043133} over Hamiltonian simulation with tensor networks \cite{banuls2013mass, PhysRevLett.112.201601, PhysRevD.101.094509, Schmoll:2023eez, PhysRevLett.132.091903} to its simulation on quantum devices in various quantum computing paradigms \cite{PhysRevX.3.041018, PhysRevA.98.032331,  PhysRevResearch.2.023015,  Shaw2020quantumalgorithms, PRXQuantum.3.020324, PRXQuantum.5.020315, PhysRevD.109.114510, araz2024statepreparationlatticefield}. 

The Schwinger model has also been investigated using truncated lightcone Hamiltonians \cite{Bergknoff:1976xr,Eller:1986nt,Yung:1991ua}. The lightcone Hamiltonian generates translations in the lightcone coordinate $x^+=(t+x)/\sqrt{2}$. An overview of lightcone quantisation is given in Reference~\cite{Anand:2020gnn} and the application of quantum computing to QFTs defined using the truncated lightcone Hamiltonian approach has been investigated in \cite{Liu:2020eoa}.

The massive Schwinger model is Quantum Electrodynamics in (1+1) dimensions describing the dynamics of fermions and photons. The Lagrangian takes the usual form,

\begin{equation}\label{eq:massiveSchwingerL}
	\mathcal{L} = -\frac{1}{4} F_{\mu\nu}F^{\mu\nu}  + \bar{\psi}\left( i \slashed\partial - g \slashed A - m\right) \psi~,
\end{equation}
where $F_{\mu\nu} \equiv \partial_\mu A_\nu - \partial_\nu A_\mu$ is the electromagnetic field tensor, $A_\mu$ is a U(1) photon field, $\psi$ is a two-component fermion field, and $g$ is the coupling strength with dimensions of mass. In (1+1) dimensions, there are no directions transverse to the momenta of moving particles. Therefore, there are no propagating photon degrees of freedom. 

To begin, we consider the massless case by setting $m=0$, such that the massless Schwinger model Lagrangian takes the form,

\begin{equation}\label{eq:masslessSchwingerL}
	\mathcal{L}_0 = -\frac{1}{4} F_{\mu\nu}F^{\mu\nu}  + \bar{\psi}\left( i \slashed\partial - g \slashed A\right) \psi~.
\end{equation}
In this limit, the model has an anomalous U(1) chiral symmetry, and the $\theta$ term can be removed with a chiral transformation. The massless Schwinger model was solved exactly by Schwinger~\cite{Schwinger:1962tp}, who showed that the model's Green's functions were those of a free massive scalar field. Therefore, it is possible to reformulate the model as a massive scalar field theory, with the Hamiltonian density~\cite{Coleman:1975pw}
\begin{equation}\label{eq:hfree}
	\mathcal{H}_0 = \frac{1}{2} : \Pi^2 + (\partial_x \phi)^2 + \frac{g^2}{\pi} \phi^2:~,
\end{equation}
where $:\,:$ denotes the normal ordering of the creation and annihilation operators used to represent the scalar field, changing the definitions of these operators, and therefore the normal ordering convention only adds an extra constant to $\mathcal{H}_0$. The mass can be read off as $M^2 \equiv g^2/\pi$, and $\Pi$ is the canonical momentum of the scalar field $\phi$. This reformulation of the model is an example of bosonisation~\cite{Coleman:1974bu,10.1093/oso/9780198834625.001.0001}. Although the Schwinger model can be bosonised, this isn't the case for generic gauge theories. However, the HT approach can still be applied in these cases by taking $H_0$ to be a solvable theory other than the free scalar field. See the discussion in Reference~\cite{fitzpatrick2022}.

In this study, we are interested in simulating the real-time evolution of interacting quantum field theories. Therefore, we will consider the massive Schwinger model from Equation~\eqref{eq:massiveSchwingerL}. The mass $m$ breaks chiral symmetry, rendering the $\theta$ term physical. The background electric field strength is related to $\theta$ through $E_B = g\theta/2\pi$~\cite{Coleman:1976uz}.

Adding the fermion mass introduces a interaction term to the bosonised Hamiltonian~\cite{Coleman:1975pw}, such that 
\begin{equation}\label{eq:hamcontinuum}
	\mathcal{H} = \mathcal{H}_0 - 2\, c\, m\, M\, :\cos \left( \sqrt{4\pi} \phi + \theta \right):~,
\end{equation}
where $\mathcal{H}_0$ is the free theory Hamiltonian from Equation~\eqref{eq:hfree}, and $c$ depends on the definition of the creation and annihilation operators which are to be normal ordered~\cite{Coleman:1974bu}.

We consider the massive Schwinger model placed on a finite circle of length $L$ (which ensures that the spectrum is discrete). On the circle, gauge fields must satisfy periodic boundary conditions, but fermion fields can satisfy $\psi(t,x) = e^{i\delta}\,\psi(t,x+L)$. Since $\delta$ can be changed using a type of gauge transformation \cite{Hetrick:1988yg}, physical quantities should not depend on this choice. After bosonisation, the scalar field then satisfies periodic boundary conditions. 

To employ Hamiltonian Truncation directly to the bosonised theory (without lattice discretisation), we decompose the Hamiltonian as in Equation~(\ref{eq:hdef}) and take as the solvable part to be
\begin{equation}\label{eq:H0}
	H_0 = \int_0^L dx \, \mathcal{H}_0 = \sum_{n=-\infty}^\infty E_n \, a^\dagger_n a_n~,
\end{equation}
where we identify the momentum $k_n = \left(2\pi n/L\right)$ for $n\in\mathbb{Z}$ and the energy $E_n=\sqrt{k^2_n+M^2}$ of the $n$\,th mode. The commutation relations for the creation and annihilation operators take the usual form: $[a_n,a_m]=[a^\dagger_n,a^\dagger_m]=0$ and $[a_n,a_m^\dagger]=\delta_{n,m}$. 

The truncated basis we use is formed from the eigenstates of Equation~\eqref{eq:H0}, which are Fock states that take the form
\begin{equation}\label{eq:basisstates}
	\ket{\{\bf r\}} = \prod_{\substack{n=-\infty}}^{n=\infty}\frac{1}{\sqrt{r_n!}}\left(a_n^\dagger\right)^{r_n}\ket{0}\,,
\end{equation}
where $\ket{0}$ is the vacuum state satisfying $a_n\ket{0}=0$ for all modes $n$.

The interaction is the integral of the second term in Eq.~(\ref{eq:hamcontinuum})
\begin{equation}\label{eq:potential}
	V = -c\,m\,M \int^L_0 dx : \exp\left[i\sqrt{4\pi} \phi(x) + i\theta \right] : + \textrm{h.c.}~,
\end{equation}
where on the finite volume circle, the scalar field $\phi$ entering Equation~\eqref{eq:potential} should be understood as the following sum over modes
\begin{equation}\label{eq:fieldDef}
	\phi(x) = \sum_{n=-\infty}^\infty \frac{1}{\sqrt{2LE_n}} \left(a_n\, e^{ik_nx} + a^\dagger_n\, e^{-ik_nx}\right)~.
\end{equation}
The sum runs over all integer mode numbers $n$, including the zero modes $n=0$ (see \cite{Schmoll:2023eez} and references therein).

We use the dots $: :$ in \eqref{eq:potential} to represent normal ordering with respect to the bosonic creation and annihilation operators of Equation~\eqref{eq:H0} and \eqref{eq:fieldDef}. For simplicity, we take the coefficient $c$ to equal its infinite volume limiting value
\begin{equation}
	c = \frac{e^\gamma}{4\pi}~,
\end{equation}
where $\gamma\approx0.57721$ is the Euler--Mascheroni constant.

Using Equation~\eqref{eq:fieldDef}, the normal-ordered exponential can be expanded in terms of the creation and annihilation operators
\begin{align}\label{eq:expandedexp}
	:\exp\left[i\sqrt{4\pi} \phi(x)\right] : \: & = \prod^\infty_{n=-\infty} \exp\left[\sqrt{\frac{2\pi}{LE_n}}\;ie^{-ik_nx}a^\dagger_n\right] \exp\left[\sqrt{\frac{2\pi}{LE_n}}\;ie^{ik_nx}a_n\right]\,,\\
	& =\prod^\infty_{n=-\infty}\sum^\infty_{j_n,j_n^\prime=0} \frac{1}{j_n^\prime ! \, j_n !} \left(i\sqrt{\frac{2\pi}{LE_n}} \right)^{j_n+j_n^\prime} e^{ik_nx\left(j_n-j_n^\prime\right)} \left(a^\dagger_n\right)^{j_n^\prime} \left(a^{\phantom{\dagger}}_n\right)^{j_n}~.
	\nonumber
\end{align}
The matrix elements of the interaction $V$ between basis states can then be calculated by combining Equation~\eqref{eq:basisstates}, \eqref{eq:potential} and \eqref{eq:expandedexp}. The result is
\begin{align}\label{eq:vfull}
	\matrixel{\{\bf r^\prime\}}{V}{\{\bf r\}}  = -cm_fMLe^{i\theta}\delta_{r_0^\prime,\,r_0}&\prod_{n=-\infty}^{n=\infty}\frac{1}{\sqrt{r^\prime_n!r_n!}}\sum_{j_n,\,j^\prime_n=0}\frac{1}{j^\prime_n!j_n!}\left(i\sqrt{\frac{2\pi}{LE_n}}\right)^{j_n+j^\prime_n} \nonumber\\ 
	&\matrixel{0}{\left(a_n^{\phantom{\dagger}}\right)^{r^\prime_n}\left(a_n^\dagger\right)^{j^\prime_n}\left(a_n^{\phantom{\dagger}}\right)^{j_n}\left(a_n^\dagger\right)^{r_n}}{0} + \text{h.c.}\,,
\end{align}
where the product of the creation and annihilation operators is given by \cite{Kukuljan:2021ddg}
\begin{equation}\label{eq:4ops}
	\matrixel{0}{\left(a_n^{\phantom{\dagger}}\right)^{r^\prime_n}\left(a_n^\dagger\right)^{j^\prime_n}\left(a_n^{\phantom{\dagger}}\right)^{j_n}\left(a_n^\dagger\right)^{r_n}}{0} = {r^\prime_n\choose j^\prime_n} {r_n\choose j_n}j^\prime_n!j_n!(r_n-j_n)!\delta_{r^\prime_n-j^\prime_n,\,r_n-j_n}\Theta(r_n-j_n)\,.
\end{equation}
The integral over space in Equation~\eqref{eq:potential} imposes momentum conservation as an additional constraint. As a result, the matrix element in Equation~\eqref{eq:vfull} vanishes unless
\begin{align}\label{eq:momcon}
	\sum_{n=-\infty}^\infty n(r_n-r^\prime_n) = 0\,,
\end{align}
where we have used the delta function in Equation~\eqref{eq:4ops} to eliminate the $j_n^{(\prime)}$ indices in favor of the occupation numbers $r_n^{(\prime)}$. In this paper, we will consider only states with vanishing total momentum.

\begin{figure}[t]
\centering
\includegraphics[width=0.8\textwidth]{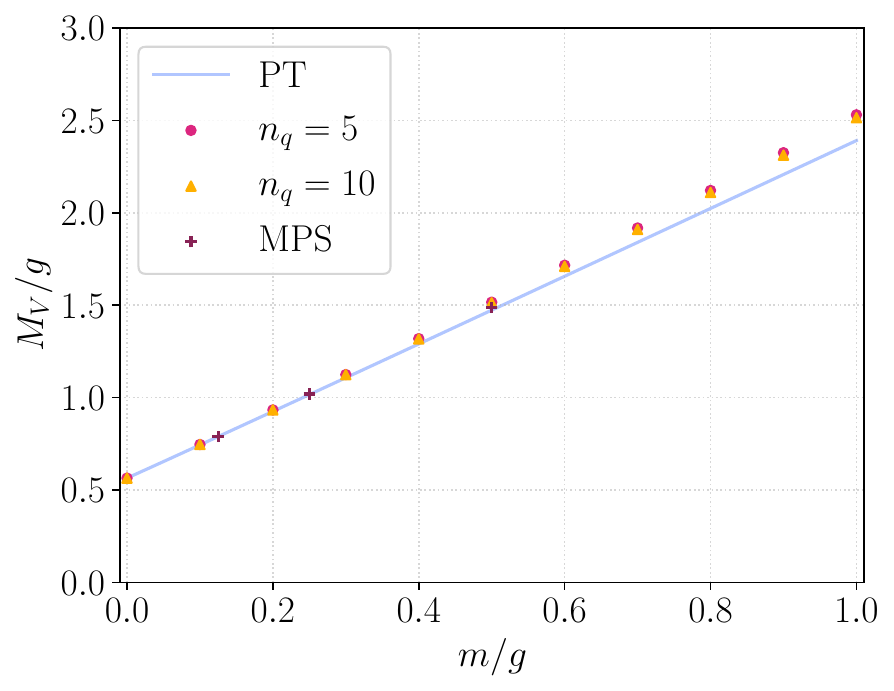}
\caption{Comparison of the vector mass, $M_V$, for varying fermion masses, $m$. The HT model shows good agreement with perturbation theory (PT) in the quantity $m/g$~\cite{Adam:1997wt} to second order, and with Matrix Product State (MPS) results taken from ~\cite{banuls2013mass}.}\label{fig:vectorMass}
\end{figure}

The final step is to build the Hamiltonian as an explicit matrix in the basis of Fock states, defined in Equation~\eqref{eq:basisstates}, with eigenvalues of $H_0$ less than, or equal to $E_\textrm{max}$. It has been explicitly shown that, for QFTs defined as conformal field theories deformed with relevant operators on a cylinder space-time, the number of basis states grows exponentially with the value of the energy cutoff, $E_\textrm{max}$~\cite{CARDY1991403, PhysRevD.91.025005}. Therefore, the size of the Hamiltonian will grow exponentially with the truncation, quickly rendering the real-time evolution of a QFT intractable on a classical device. A qubit-based quantum computer's exponentially growing Hilbert space allows for efficient information encoding. Therefore, if the size of the Hamiltonian grows exponentially with $E_\textrm{max}$, then the scaling of the number of required qubits will grow at most polynomially. 

For the analysis presented in this paper, we will only take $\theta = 0$, in which case the Schwinger model has two additional discrete symmetries: Parity (P) and Charge Conjugation (C). In the bosonised theory, the action of C is to send $\phi(x,t)\rightarrow-\phi(x,t)$, and P sends $\phi(x,t)\rightarrow-\phi(-x,t)$. We therefore divide our truncated Hilbert space into four subsectors of states which are even or odd under the two symmetries, and construct separate smaller truncated Hamiltonians in the relevant subsectors, thus simplifying our analysis.

For quantum computing applications, it is convenient to redefine the cutoff of the truncated Hilbert space using the number of qubits, $n_q$, required to represent time evolution using the Hamiltonian (within a particular symmetry subsector). To do this, we order the relevant states in energy and take the first $2^{n_q}$ as our truncated basis, such that the Hamiltonian has the size $\left( 2^{n_q} \times 2^{n_q} \right)$.

To test our framework, we first consider the value of the vector mass in the Schwinger model, at different fermion mass values. This mass is simply the difference in energy between the lowest energy states in the C even P even, and the C odd P odd subsectors. Figure~\ref{fig:vectorMass} shows a comparison of the vector mass calculated using the HT approach compared with calculations from second order perturbation theory~\cite{Adam:1997wt} and Matrix Product States (MPS)~\cite{banuls2013mass}. To construct the HT calculation we choose the volume $(g \,L) =8$.
Since finite volume effects are exponentially suppressed for $M_VL\gg1$~\cite{Lüscher1984}, we expect finite volume effects to be small with this choice.

In Figure~\ref{fig:vectorMass}, we see that the HT method is well converged even for small truncation values, agreeing well with both perturbative and non-perturbative methods at small fermion masses with a truncation of $n_q=5$. The comparisons between HT and the two infinite-volume methods also confirm that finite volume effects are small for $(g \,L) =8$. We see exact agreement between all methods at a value of $(m/g)=0.2$, and this value will be used in Section~\ref{sec:QC} when simulating the time-evolution of the system on a quantum device. 

We note that while energy differences between states are well converged, the ground state energy diverges as $\sim m^2\log E_\textrm{max}$, just like the free massless fermion theory when perturbed with a mass term in 1+1d. Divergences that affect individual state energies but not energy differences do not affect time-evolution, or the determination of particle masses and can be renormalised away in HT by introducing an effective Hamiltonian~\cite{EliasMiro:2022pua,Delouche:2023wsl}.


\section{Truncated Schwinger Model on a Quantum Device} \label{sec:QC}

In the Hamiltonian framework, the real-time evolution of the quantum system,

\begin{equation}
U(t)~=~e^{-iHt} \,, 
\end{equation}
can be naturally implemented on a quantum device by using a product formula method based on the Trotter-Suzuki decomposition~\cite{MR0103420, 10.1063/1.526596, PhysRevX.11.011020}. Consider a Hamiltonian expressed as a sum of non-commuting operators, $H = \sum_i H_i$. The time-evolution operator, $U(t)$, can be approximated by 

\begin{equation}\label{eq:trotter}
\mathcal{U}(t) = \left[ \prod_i e^{-i H_i t/n} \right]^n~,
\end{equation}
up to an error $\mathcal{O}(t^2/n)$, where $n$ is a positive integer. The operator $\mathcal{U}(t)$ defines the Trotterised time-evolution, which divides the total evolution time, $t$, into $n$ steps of time $\delta t = t/n$. The total time-evolution is therefore achieved by iteratively applying $n$ so-called Trotter steps, such that the Trotterised time-evolution is exact in the limit $n\rightarrow \infty$. 

In this Section, we demonstrate how the Trotterisation method can be used to simulate the real-time evolution of the Schwinger model constructed from the Hamiltonian Truncation (HT) approach on a quantum device. We test the model's suitability for Noisy-Intermediate Scale Quantum (NISQ) devices by evaluating the algorithm's performance at different truncations and ultimately running the simulation on the \texttt{ibm\_brisbane} quantum computer, which operates a 127-qubit Eagle R3 processor. 

\subsection{Time-evolution via quantum simulation} 

To efficiently simulate the time evolution of a quantum field theory (QFT) on a quantum device, the Hamiltonian of the model must first be mapped onto a basis corresponding to operations native to the quantum device. For a qubit-based quantum computer, such as the \texttt{ibm\_brisbane} device used for this paper, a suitable basis of operations is constructed from tensor products of the Pauli operators and an identity operator, $\left(\sigma_0, \sigma_1, \sigma_2, \sigma_3 \right)$, such that

\begin{equation}
H = \sum_{{i_1}, \hdots, {i_{n_q}}=0}^3 \alpha_{{i_1}, \hdots, {i_{{n_q}}}} \left( \sigma_{{i_1}} \otimes \hdots \otimes \sigma_{{i_{n_q}}} \right)~,
\end{equation}
where $\alpha_{{i_1}, \hdots, {i_{{n_q}}}}$ is the coefficient of the corresponding Pauli term, $( \sigma_{{i_1}} \otimes \hdots \otimes \sigma_{{i_{n_q}}})$. The exponentials of Pauli terms can be implemented on a qubit-based quantum device through a sequence of single-qubit rotation gates and \textsc{cnot} gates. A circuit can then be constructed for a single Trotter step of the time-evolution by calculating the form of Equation~\eqref{eq:trotter}. The number of qubits required and the circuit depth for a single Trotter step will grow with the size of Hamiltonian. The time-evolution of the QFT is then simulated on a quantum device by following three steps: (1) prepare the initial state, (2) apply the Trotter step circuit iteratively for $n$-Trotter-steps, and (3) measure the circuit with respect to an observable. 

Preparing the initial state is a highly non-trivial task and has been shown to require exponential circuit depths to construct arbitrary quantum states without ancillary qubits~\cite{Xiaoming10044235}. Using ancillary qubits, the circuit depth scaling can be reduced to polynomial scaling at the cost of an exponentially growing number of ancillary qubits~\cite{PhysRevA.83.032302, PhysRevResearch.3.043200, PhysRevLett.129.230504}. The final step in the time evolution also requires a state-preparation circuit to rotate the system into a basis such that the desired observable can be measured. Therefore, the circuit depth can grow very rapidly due to the state preparation schemes that need to be implemented, and this has been identified as one of the limiting factors of implementing Lattice models on a NISQ device. 

In contrast, the HT method allows for the Hamiltonian to be constructed such that the ground state of the free theory, $H_0$, corresponds exactly to the ground state of the qubit-based quantum device, namely the zeroth state in the computational basis. As a result, complicated and costly state-preparation routines are not required to prepare the ground state of the system. This approach, therefore, offers an advantage over Lattice models for simulating QFTs on NISQ devices by reducing the circuit depth requirements for time evolution.

The size of the Hamiltonian, and by extension, the number of qubits required to implement the time-evolution, of the Schwinger model obtained from HT is determined by the cutoff energy, $E_\textrm{max}$, and the volume, $L$. In this paper, we will vary the cutoff energy only, setting the volume, $(g\, L)=8$, at a large enough value to approximate the theory well. From Equation~\eqref{eq:hfree}, the scalar mass is $M=g/\sqrt{\pi}$. The Hilbert space is bosonic and is constructed from the eigenbasis of the free Hamiltonian in Equation~\eqref{eq:hfree}. As a result, it is not trivial to represent the operators within the Hamiltonian in terms of the fermion creation and annihilation operators. Thus, the massive Schwinger model Hamiltonian from Equation~\eqref{eq:hamcontinuum} cannot be easily decomposed into tensor products of Pauli operators, which are natural operations for qubit based devices. For this model, this may be a limiting factor for the HT approach as the number of qubits increases because Pauli decomposition leads to an exponentially growing number of Pauli terms in the decomposition of the Hamiltonian. New techniques in the construction of the Hilbert space may be needed to achieve the true potential of this approach. In this paper, it will be shown that systems with small numbers of qubits are sufficient to describe the dynamics of the model and that the approach is well suited to NISQ devices. 

An interesting process to study is the post-quench dynamics of the Schwinger model.  A quench is achieved by preparing the system in the ground state of the free theory, $H_0$, and then instantaneously ``switching on" the potential term, $V$, such that the system now evolves under the total Hamiltonian, $H = H_0 + V$. For the Schwinger model presented in Section~\ref{sec:HT}, this is equivalent to switching on the fermion mass, $m$. To describe this process, we only need to construct and use the Hamiltonian in the C even, P even subsector, since our initial state lies in this subsector, and the C and P symmetries prevent time evolution to states outside this subsector. The state of the system after post-quench evolution is, therefore,

\begin{equation}
\left\vert \psi (t) \right\rangle = e^{-i H t} \left\vert \psi (0) \right\rangle\approx  \left[ \prod_i e^{-i H_i t/n} \right]^n \left\vert \psi(0) \right\rangle~.
\end{equation}
The dynamics of the model can then be examined by establishing the probability that the system is in the ground state of the free, $m=0$ theory. We define the time-dependent observable $G(t)$, such that 

\begin{equation}
G(t) = \left\langle \textrm{vac} \left\vert e^{-i H t} \right\vert \textrm{vac} \right\rangle~,
\end{equation}
where the ground state of the free Hamiltonian is defined as $\left\vert \textrm{vac} \right\rangle = \left\vert 0 \right\rangle$ in the computational basis of the qubit device. We will examine the time-dependence of $\left|G(t)\right|^2$, and explore the effects of truncation and Trotterisation errors on this probability, and investigate the feasibility of simulating the model on NISQ devices.  

\begin{figure}[t]
\centering
\includegraphics[width=\textwidth]{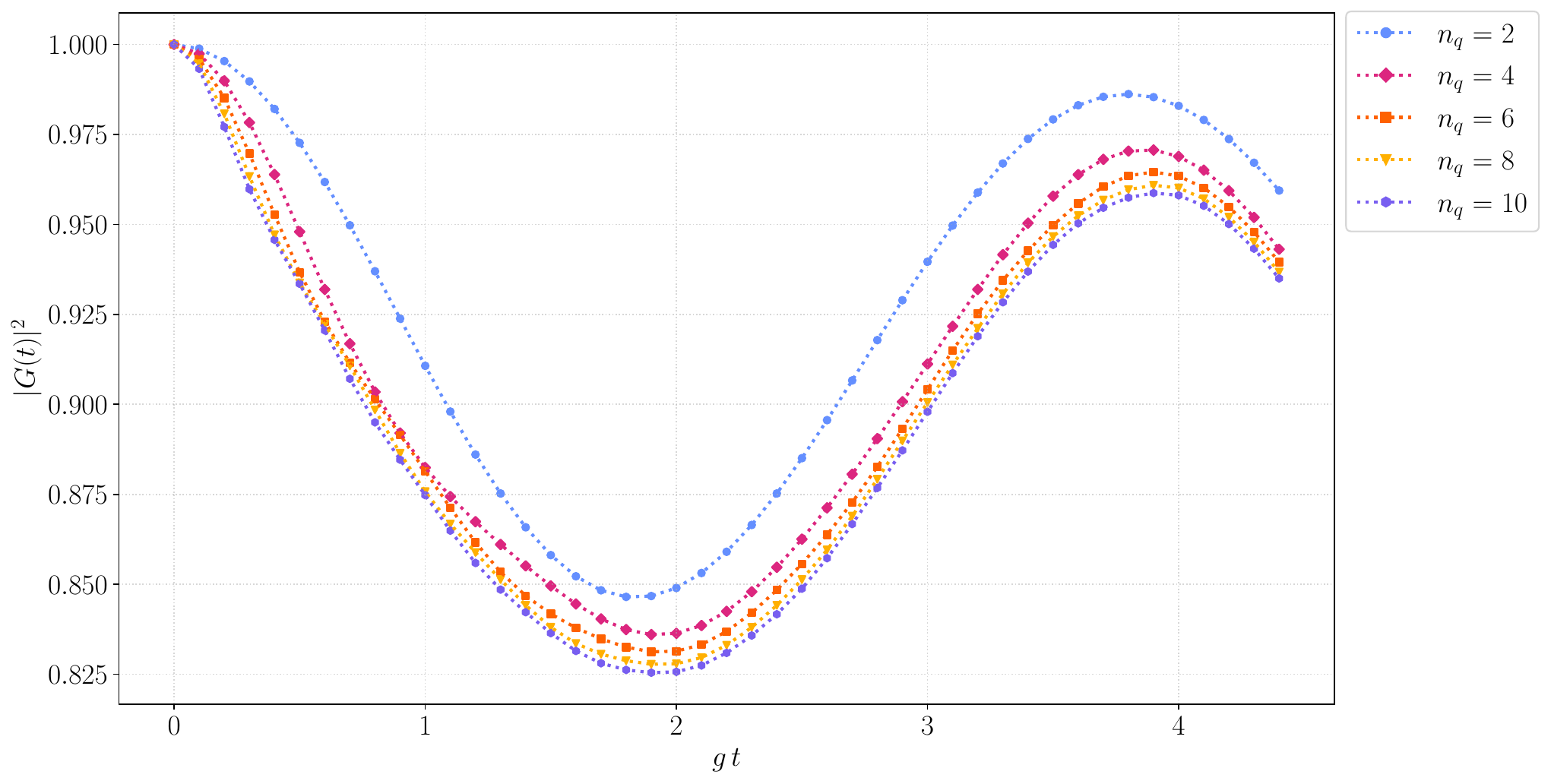}
\caption{Comparison of the time-evolution of the Schwinger model at different truncations with $(m/g)=0.2$. The time-evolution has been calculated by the Exp-method: brute-force exponentiation of the Hamiltonian. The HT method shows quick convergence as the truncation increases, allowing for the dynamics to be reliably captured using only a small number of qubits.}\label{fig:truncationCheck}
\end{figure}

It is important to quantify the error introduced by the truncation by comparing the accuracy of the time-evolution calculation at different truncations. In practice, the task is to achieve a trade-off between the resources required to simulate the model and the truncation error, minimising both as much as possible. Figure~\ref{fig:truncationCheck} shows one period of the time-evolution of the quenched system from Equation~\eqref{eq:hamcontinuum} with $(m/g)=0.2$ for different truncations. Here, the time evolution has been calculated by brute-force exponentiation of the Hamiltonian in a method we will call the \textit{Exp}-method. This method simulates the time evolution without Trotterisation errors. The model exhibits good convergence as the truncation increases and performs remarkably well for small truncations, describing the post-quench dynamics well. The good convergence we see validates our parameter choice $(m/g)=0.2$.

\begin{figure}[t]
\centering
\includegraphics[width=\textwidth]{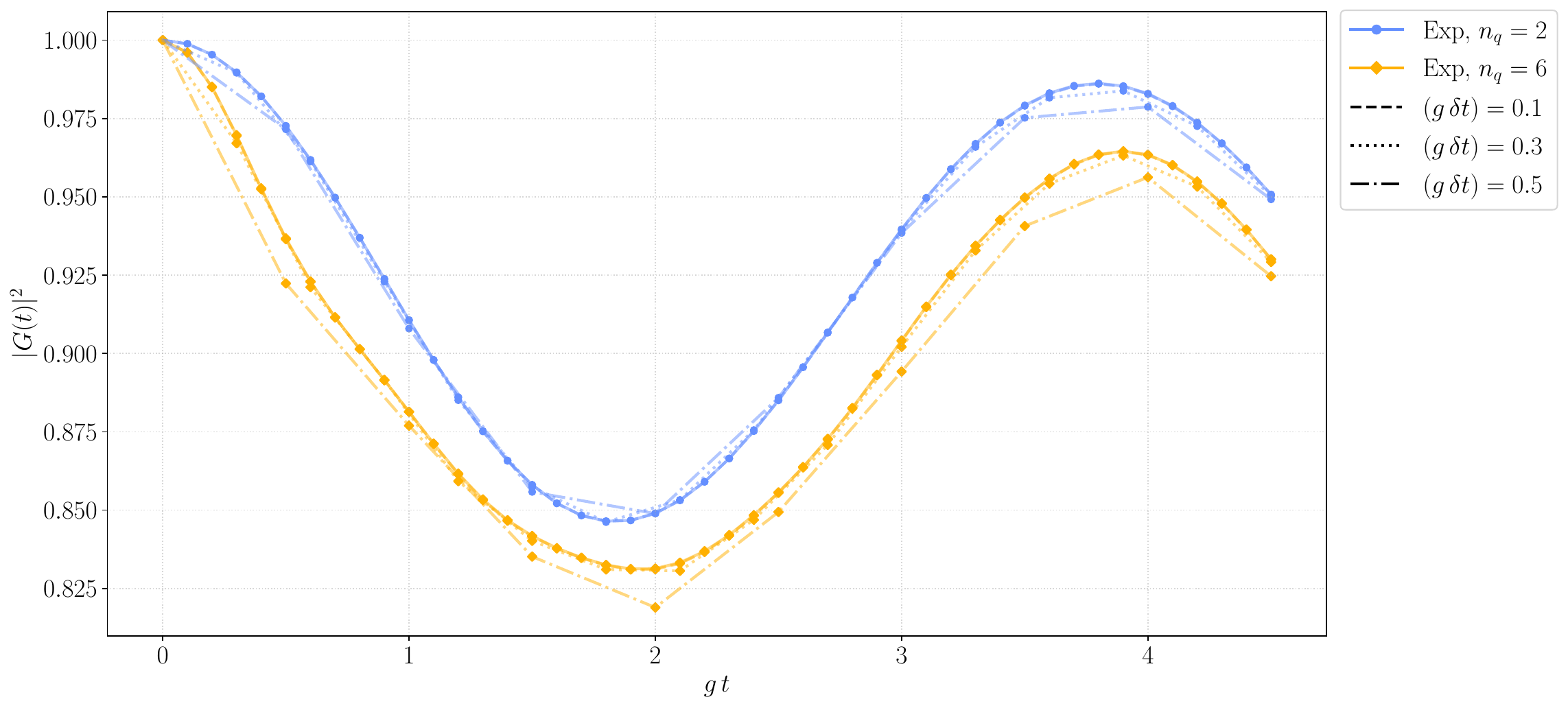}
\caption{Time-evolution of the Schwinger model with varying Trotter time-steps compared to the brute-force exponentiation of the Hamiltonian, so-called Exp. The Trotterised time-evolution was performed on a quantum simulator that simulates a fully fault-tolerant quantum device for truncations of $n_q = 2$ (blue) and $n_q=6$ (yellow). The system exhibits increasing Trotter errors when taking more time steps.}\label{fig:trotterCheck}
\end{figure}

Together with the truncation, another significant source of error is the so-called ``Trotter error" arising from approximating the time evolution operator from Equation~\eqref{eq:trotter}. This study will use first-order Trotterisation, which approximates the time-evolution up to an error $\mathcal{O}\left( t^2/n \right)$. Therefore, to minimise the error induced by Trotterisation, one must split the time-evolution into sufficiently small time steps, $\delta t$. However, performing many Trotter steps increases the resource cost of simulating the time evolution, specifically increasing the required circuit depth for quantum simulations. Therefore, again, there is a trade-off between choosing a small enough Trotter step and the resources used. Figure~\ref{fig:trotterCheck} shows the time-evolution of the Schwinger model for different values of $(g\,\delta t)$ compared to the Exp-method. The simulations have been executed with a truncation of $n_q=2$ and $n_q=6$, as shown by the blue and yellow lines in Figure~\ref{fig:trotterCheck} respectively. The Trotterised time-evolution has been performed using the \texttt{BaseSampler} quantum simulator from \textsc{Qiskit}~\cite{qiskit2024}, which simulates a fully fault-tolerant quantum device. We see that the model is remarkably resilient to Trotter error and that the time-evolution with $(g\,\delta t)=0.1$ agrees exactly with the Exp-method, even at larger evolution times. However, as the Trotter step increases, the error is more pronounced at large evolution times. This indicates a Trotter error, which grows cumulatively with each Trotter step. As the time step is increased to $(g\,\delta t) = 0.5$, it is clear that the Trotter error increases, and the distribution deviates away from the Exp-method more quickly for both cases. 

Mitigating theoretical errors whilst also remaining within realistic resource constraints of NISQ devices is a challenge for all quantum computing approaches to simulating QFTs. From Figure~\ref{fig:truncationCheck}, we see that the HT method allows for good convergence without needing many qubits. Similarly, we see in Figure~\ref{fig:trotterCheck} that the time-evolution of the Schwinger model via HT does not suffer greatly from Trotter error, even for relatively large Trotter time steps. For this reason, quantum simulation via HT is potentially well suited to NISQ devices, which have few qubits and short decoherence times. 
 
\subsection{Schwinger model on NISQ devices}\label{sec:QCResults}

To determine the feasibility of using the HT approach to simulating QFTs on a quantum device, the time-evolution of the Schwinger model has been run on the $\texttt{ibm\_brisbane}$ quantum computer, a 127-qubit device with an Eagle R3 quantum processor. The model has been constructed to use minimal resources whilst retaining the interesting dynamics of the post-quench system. Therefore, the Hamiltonian from Equation~\eqref{eq:hamcontinuum} has been truncated to run on $n_q=2$ qubits with a $(m/g)=0.2$ coupling. A Trotter time-step of $(g\,\delta t) =0.3$ has been chosen to limit the algorithm's circuit depth and mitigate decoherence. To further suppress the errors from the quantum computer, the AI-driven error suppression pipeline \textsc{Q-Ctrl}~\cite{PhysRevApplied.15.064054, boulder_opal1, boulder_opal2} has been used through the \textsc{Qiskit} platform~\cite{qiskit2024}. 

\begin{figure}[t]
\centering
\includegraphics[width=\textwidth]{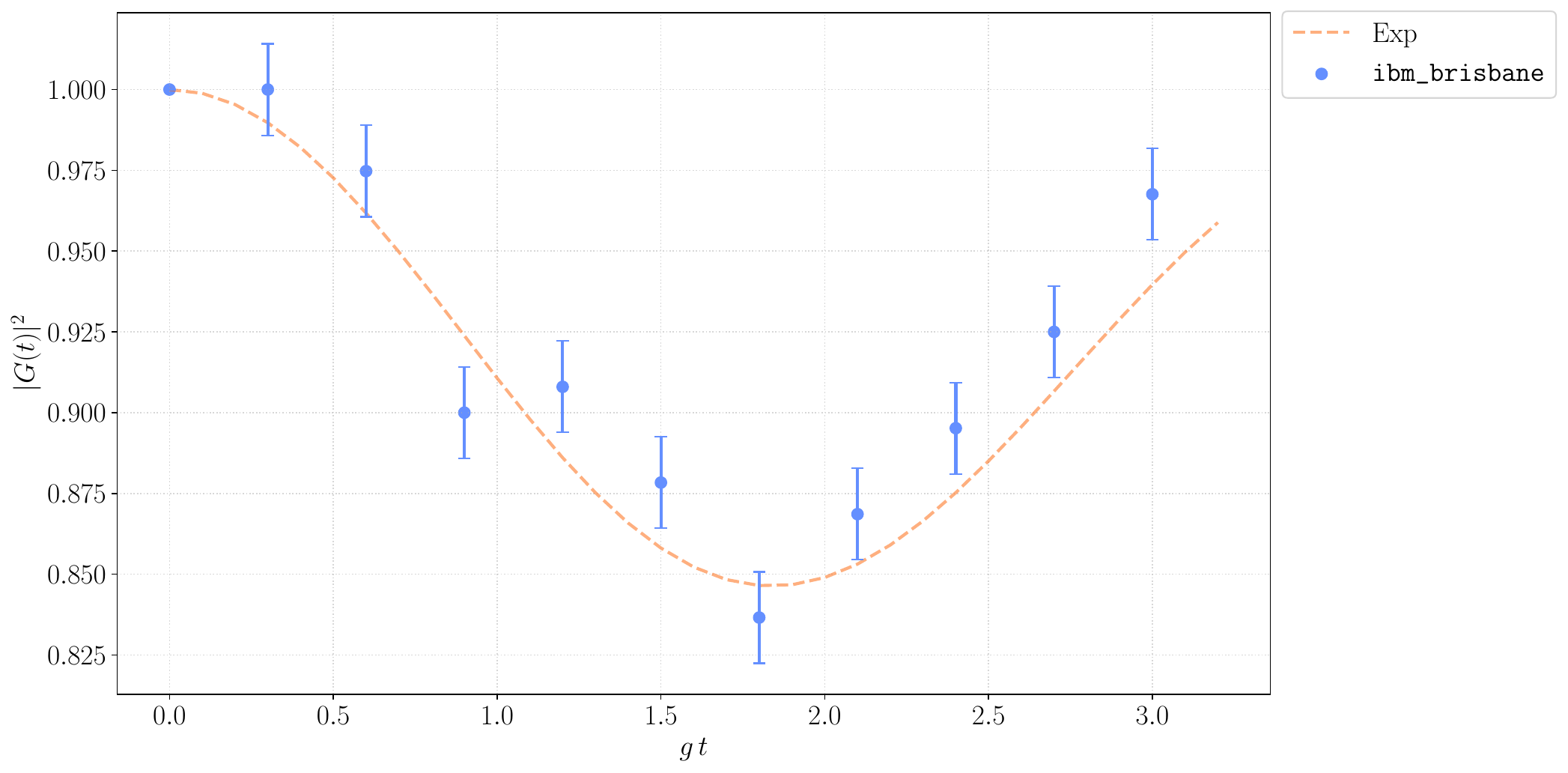}
\caption{Time-evolution of the Schwinger model via HT run on the \texttt{ibm\_brisbane} 127-qubit quantum computer. The results have been enhanced using error mitigation and suppression routines through \textsc{Qiskit}~\cite{qiskit2024} and \textsc{Q-Ctrl}~\cite{boulder_opal1, boulder_opal2}. The output of the quantum device agrees well with the brute-force exponentiation of the Hamiltonian, Exp, demonstrating the suitability of the HT approach to the simulation of QFTs on NISQ devices.}\label{fig:QCrun}
\end{figure}

Figure~\ref{fig:QCrun} presents the results from the quantum computer compared to the Exp-method. Each data point has been generated from 1000 shots on the quantum computer, and the statistical errors are displayed. The quantum device's performance is remarkable, showing strong agreement with the Exp-method and closely matching the simulator's performance for $(g\,\delta t) = 0.3$, as shown in Figure~\ref{fig:trotterCheck}. These results highlight the suitability of the HT approach to quantum simulation for NISQ devices. With the interesting physics of the Schwinger model's post-quench dynamics being captured using only a small number of qubits, and without the need for complicated and costly state preparation, HT provides a promising route towards simulating the time-evolution of QFTs on near-term quantum devices.


\section{Conclusion}

This study demonstrates the viability of using Hamiltonian Truncation (HT) techniques to facilitate the non-perturbative, real-time simulation of Quantum Field Theories (QFTs) on Noisy-Intermediate Scale Quantum (NISQ) devices. By focusing on the Schwinger model, a (1+1) dimensional Quantum Electrodynamics system, we have showcased our approach's practicality and benefits.

Our findings indicate that HT significantly reduces the complexity of the problem by removing the need for complicated and costly state preparation routines, reducing the overall circuit depth of the algorithm. Furthermore, for the observables studied in this paper, we have shown that the HT approach converges quickly, thus capturing the dynamics of the system without requiring many qubits. Combined, these attributes of HT make it feasible to implement on NISQ devices, which have limited qubits and coherence times. 

We demonstrate the suitability of the HT approach for NISQ devices by running a time-evolution algorithm on the \texttt{ibm\_brisbane} quantum computer using only two qubits. We employed the Trotter-Suzuki decomposition to approximate the time-evolution operator, facilitating the implementation of real-time dynamics on the quantum device. Our results show good agreement with both the classical brute-force exponentiation of the Hamiltonian, the Exp-method in Figure~\ref{fig:QCrun}, even when executed on hardware with inherent noise and operational constraints. Despite the small number qubits, the system gives a good qualitative description of the underlying physics. 

In conclusion, our work highlights the potential of HT as a viable pathway for simulating QFTs on quantum hardware. It encourages more complex simulations and offers a scalable and efficient approach for leveraging quantum devices in non-perturbative field theory research. Future work will extend this approach to more complex models and explore its applicability to a broader range of quantum simulations.

\vspace{0.5cm}
\noindent{\textit{\textbf{Acknowledgments} We acknowledge the use of IBM Quantum services for this work. The views expressed are those of the authors, and do not reflect the official policy or position of IBM or the IBM Quantum team.} 

\bibliographystyle{inspire}
\bibliography{refs}{}

\end{document}